
%
%
\font\straight=cmss10
\font\tenbifull=cmmib10 \skewchar\tenbifull='177
\font\tenbimed=cmmib7   \skewchar\tenbimed='177
\font\tenbismall=cmmib5  \skewchar\tenbismall='177
\textfont9=\tenbifull
\scriptfont9=\tenbimed
\scriptscriptfont9=\tenbismall
\def\bmit{\fam9}
\mathchardef\alpha="710B
\mathchardef\beta="710C
\mathchardef\gamma="710D
\mathchardef\delta="710E
\mathchardef\epsilon="710F
\mathchardef\zeta="7110
\mathchardef\eta="7111
\mathchardef\theta="7112
\mathchardef\iota="7113
\mathchardef\kappa="7114
\mathchardef\lambda="7115
\mathchardef\mu="7116
\mathchardef\nu="7117
\mathchardef\micron="716F
\mathchardef\xi="7118
\mathchardef\pi="7119
\mathchardef\rho="711A
\mathchardef\sigma="711B
\mathchardef\tau="711C
\mathchardef\upsilon="711D
\mathchardef\phi="711E
\mathchardef\chi="711F
\mathchardef\psi="7120
\mathchardef\omega="7121
\mathchardef\varepsilon="7122
\mathchardef\vartheta="7123
\mathchardef\varphi="7124
\mathchardef\varrho="7125
\mathchardef\varsigma="7126
\mathchardef\varpi="7127
%
\magnification=1200
\baselineskip=13pt
\overfullrule=0pt
\tolerance=100000
\def\pr{\hbox{\straight pr}}
\def\v{\hbox{\straight v}}
\def\qb{q^*}
\def\thetab{\theta^*}
{\hfill \hbox{\vbox{\settabs 1\columns
\+ UR-1391 \cr
\+ ER-40685-840\cr
\+ hep-th/9410165\cr
}}}

\bigskip

\baselineskip=18pt

\centerline{\bf KdV and NLS Equations as Tri-Hamiltonian Systems}

\vfill

\centerline{J. C. Brunelli}
\medskip
\centerline{and}
\medskip
\centerline{Ashok Das}
\medskip
\medskip
\centerline{Department of Physics and Astronomy}
\centerline{University of Rochester}
\centerline{Rochester, NY 14627, USA}
\vfill

\centerline{\bf {Abstract}}

\medskip

We show that the KdV and the NLS equations are tri-Hamiltonian systems. We
obtain the third Hamiltonian structure for these systems and prove Jacobi
identity through the method of prolongation. The compatibility of the
Hamiltonian structures is verified directly through prolongation
as well as through
the shifting of the variables. We comment on the properties of the recursion
operator as well as the connection with the two boson hierarchy.
\vfill
\eject

\noindent {\bf 1. {Introduction}}

\smallskip

Most integrable models in $0+1$, $1+1$ and $2+1$ dimensions are known to be
bi-Hamiltonian systems [1-5]. These are systems whose dynamical equations can
be
described through Hamilton's equations with respect to two distinct Hamiltonian
structures  which are also compatible [5], namely,
any linear superposition of the
two also defines a Hamiltonian structure. Since the Jacobi identity involves a
nonlinear relation, compatibility of Hamiltonian structures is a nontrivial
statement. There is only one known $1+1$ dimensional integrable system, namely,
the two boson hierarchy or the equation describing long water waves [6,7-12],
which is
even a tri-Hamiltonian system. Namely, the dynamical equations for this system
can be written in the Hamiltonian form with respect to three distinct
Hamiltonian structures which are compatible in the sense that any arbitrary,
linear superposition of three of them is also a Hamiltonian structure [13].

It is quite surprising that the two boson hierarchy is the only
known integrable system  which is
tri-Hamiltonian. This result is even more surprising considering the fact that
several other integrable systems can be embedded into this system [6,7,10,12]
and yet they
are only bi-Hamiltonian. This motivated us to examine two of the most familiar
integrable systems -- the KdV equation and the nonlinear Schr\"odinger (NLS)
equation -- in detail and show that these systems are tri-Hamiltonian as well.
The third Hamiltonian structures for these systems
are highly nontrivial and we use the method of
prolongation [14] to verify the Jacobi identity as well as the compatibility
conditions. The paper is organized as follows. In sec. 2, we derive the third
Hamiltonian structure  for the KdV equation, prove the Jacobi identity and
compatibility. In sec. 3, we construct the third Hamiltonian structure for
the NLS equations and show that it is a tri-Hamiltonian system as well.
In sec. 4, we
construct the three Hamiltonian structures associated with the two boson
hierarchy starting from the NLS equation and present a brief conclusion in sec.
5. In the appendix, we compile a list of formulae for prolongation which are
useful in checking various identities.
\medskip

\noindent {\bf 2. {KdV as a Tri-Hamiltonian System}}

\smallskip

It is well known that the KdV (Korteweg-de Vries) equation
$$
{\partial u\over\partial t}=u{\partial u\over\partial x}+{\partial^3
u\over\partial x^3}\eqno(2.1)
$$
is a bi-Hamiltonian system [5]. Namely, let (at equal times)
$$
\{u(x),u(y)\}_1={\cal D}_1\delta(x-y)
={\partial\ \over\partial x}\delta(x-y)\eqno(2.2)
$$
and
$$
H_3=\int dx\,\left({1\over 3!}u^3-{1\over 2}\left({\partial u\over\partial x}
\right)^2\right)\eqno(2.3)
$$
Then, it is easily verified that
$$
{\partial u\over\partial t}=\{u(x),H_3\}_1=u{\partial u\over\partial x}+
{\partial^3 u\over\partial x^3}\eqno(2.4)
$$
showing that the KdV equation is Hamiltonian. The anti-symmetry of the
Hamiltonian structure, ${\cal D}_1$, is obvious and Jacobi identity is
trivially satisfied since this is a constant structure (independent of
dynamical variables.). (In modern terminology, one would say that the relation
(2.2) describes the U(1) current algebra with $u(x)$ considered as a current.)

We also note that if we define
$$
\{u(x),u(y)\}_2={\cal D}_2\delta(x-y)=\left({\partial^3\ \over\partial
x^3}+{1\over3}\left({\partial\ \over\partial x}u(x)
+u(x){\partial\ \over\partial x}\right)\right)\delta(x-y)\eqno(2.5)
$$
and
$$
H_2=\int dx\,{1\over 2}u^2\eqno(2.6)
$$
then, we can again write
$$
{\partial u\over\partial t}=\{u(x),H_2\}_2=u{\partial u\over\partial x}+
{\partial^3 u\over\partial x^3}\eqno(2.7)
$$
Namely, the KdV equation is also Hamiltonian with respect to a distinct, second
Hamiltonian structure. That ${\cal D}_2$ defines a Hamiltonian structure can be
seen as follows. First, the antisymmetry of ${\cal D}_2$ is obvious from the
definition in Eq. (2.5). However, because the structure now depends on the
dynamical variables, Jacobi identity is no longer automatic. On the other hand,
we recognize Eq. (2.5) as defining the Virasoro algebra [15] (think
of $u(x)$ as the
energy-momentum tensor) and, therefore, Jacobi identity must hold.
Compatibility now follows from the simple observation that ${\cal D}_2(u)$ is
Hamiltonian for any field variable $u$ satisfying Eq. (2.5) and, therefore,
${\cal D}_2(u+{3\over2}\lambda)$ where $\lambda$ is an arbitrary constant, must
also define a Hamiltonian structure. But, by definition (see Eq. (2.5))
$$
{\cal D}_2(u+{3\over2}\lambda)={\cal D}_2(u)+\lambda{\cal D}_1\eqno(2.8)
$$
and since an arbitrary linear combination of ${\cal D}_1$ and ${\cal D}_2$
defines a Hamiltonian structure, they are compatible. (This result on
compatibility can be directly verified as well.)

Let us next note that we can define
$$
\eqalign{
\{u(x),u(y)\}_3={\cal D}_3\delta(x-y)
=\Bigl(\partial^5&+{1\over3}(\partial^3u+\partial^2u\partial+\partial
u\partial^2+u\partial^3)\cr
&+{1\over9}(\partial u^2+u\partial u+u^2\partial+
\partial u\partial^{-1}u\partial)\Bigr)\delta(x-y)\cr
}
\eqno(2.9)
$$
with
$$
\partial\equiv {\partial\ \over\partial x}\eqno(2.10)
$$
and
$$
H_1=3\int dx\,u\eqno(2.11)
$$
to obtain
$$
{\partial u\over\partial t}=\{u(x),H_1\}_3=u{\partial u\over\partial x}+
{\partial^3 u\over\partial x^3}\eqno(2.12)
$$
Thus, if we can show that ${\cal D}_3$ in Eq. (2.9) has the necessary
antisymmetry property and satisfies the Jacobi identity, then this would define
a third Hamiltonian structure of the KdV equation.

The antisymmetry of ${\cal D}_3$ is obvious from the definition in Eq. (2.9).
Jacobi identity is normally easier to check by examining the closure of the
corresponding symplectic form. However, we note that the structure of ${\cal
D}_3$ is highly nontrivial, making it extremely difficult to invert. Thus, we
will check Jacobi identity for the Hamiltonian structure, ${\cal D}_3$,
directly, using the method of prolongation. We refer the interested reader to
ref.14 (see chapter 7) for details on this method and simply note
that in the infinite dimensional space labelled
by $(u,u_x,u_{xx},u_{xxx},\dots)$ if we define a bivector
$$
{\bmit\Theta}_{{\cal D}_3}={1\over2}\int dx\, {\theta}
\wedge{\cal D}_3 {\theta}\eqno(2.13)
$$
then ${\cal D}_3$ would satisfy the Jacobi identity provided
$$
\pr\,\v_{{\cal D}_3{\theta}}({\bmit\Theta}_{{\cal D}_3})=0\eqno(2.14)
$$
Here the assumption is that
$$
\theta\ne\theta[u]\eqno(2.15)
$$
and by definition prolongation acts only on coefficients functionally dependent
on $u$.

For the structure ${\cal D}_3$ in Eq. (2.9), we note that (The subscript $x$
denotes a derivative with respect to $x$.)
$$
\displaylines{
\qquad{\bmit\Theta}_{{\cal D}_3}={1\over 2}\int dx\,
\Biggl\{\theta\wedge\theta_{xxxxx}+{2\over3}u\theta\wedge\theta_{xxx}
-{2\over3}u\theta_x\wedge\theta_{xx}\hfill\cr
\hfill+{1\over3}u^2\theta\wedge\theta_x
-{1\over9}u\theta_x\wedge(\partial^{-1}u\theta_x)
\Biggr\}\qquad(2.16)\cr
}
$$
which leads to
$$
\displaylines{
\quad\pr\,\v_{{\cal D}_3{\theta}}({\bmit\Theta}_{{\cal D}_3})=
{1\over2}\int dx\,\Biggl\{{2\over3}\pr\,\v_{{\cal D}_3{\theta}}(u)
\wedge(\theta\wedge\theta_{xxx}-\theta_x\wedge\theta_{xx}
+u\theta\wedge\theta_x)\hfill\cr
\hfill-{1\over 9} \pr\,\v_{{\cal D}_3{\theta}}(u)\wedge\theta_x
\wedge(\partial^{-1}u\theta_x)
+{1\over 9}u\theta_x\wedge\left(\partial^{-1}
\pr\,\v_{{\cal D}_3{\theta}}(u)\wedge\theta_x\right)\Biggr\}
\quad(2.17)\cr
}
$$
With
$$
\displaylines{
\quad\pr\,\v_{{\cal D}_3{\theta}}(u)=\theta_{xxxxx}+
{1\over3}(u\theta)_{xxx}
+{1\over3}(u\theta_x)_{xx}+{1\over3}(u\theta_{xx})_{x}+
{1\over3}u\theta_{xxx}\hfill\cr
\hfill+{1\over9}(u^2\theta)_x +{1\over9}u(u\theta)_x+{1\over9}u^2\theta_x
+{1\over9}\left(u(\partial^{-1}u\theta_x)\right)_x\quad(2.18)\cr
}
$$
it is tedious but straightforward to show that
$$
\pr\,\v_{{\cal D}_3{\theta}}({\bmit\Theta}_{{\cal D}_3})=0\eqno(2.19)
$$
This proves that ${\cal D}_3$ in Eq. (2.9) satisfies the Jacobi identity and,
therefore, defines a third Hamiltonian structure for the KdV equation.

To prove that the three Hamiltonian structures are compatible, we define
$$
{\cal D}={\cal D}_3+\alpha{\cal D}_2+\beta{\cal D}_1\eqno(2.20)
$$
where $\alpha$ and $\beta$ are arbitrary, independent, constant parameters. By
construction $\cal D$ is antisymmetric since the three Hamiltonian structures
are. If we now construct the bivector
$$
{\bmit\Theta}_{{\cal D}}={1\over2}\int dx\,{\theta}
\wedge{\cal D}{\theta}
={1\over2}\int dx\,
{\theta}\wedge({\cal D}_3\theta+\alpha{\cal D}_2\theta+\beta{\cal D}_1
\theta)\eqno(2.21)
$$
then, once again, it is straightforward to show that (We list the formulae for
prolongation in the appendix.)
$$
\pr\,\v_{{\cal D}{\theta}}({\bmit\Theta}_{{\cal D}})=0\eqno(2.22)
$$
This shows that $\cal D$ satisfies the Jacobi identity and consequently is a
genuine Hamiltonian structure for arbitrary and independent $\alpha$ and
$\beta$. Therefore, the three Hamiltonian structures of the KdV equation are
compatible making it a tri-Hamiltonian system much like the two boson
hierarchy [6].

We note here that the compatibility of the Hamiltonian structures can be seen
alternately by shifting the dynamical variable as follows. Note that ${\cal
D}_3(u)$ defines a Hamiltonian structure for any variable $u$ satisfying the
Poisson bracket relation in Eq. (2.9). In particular, if we let
$$
u\to u+{3\over2}\lambda\eqno(2.23)
$$
where $\lambda$ is an arbitrary constant, ${\cal D}_3(u+{3\over2}\lambda)$
defines a hamiltonian structure. On the other hand,
$$
{\cal D}_3(u+{3\over2}\lambda)=
{\cal D}_3(u)+2\lambda{\cal D}_2(u)+\lambda^2{\cal
D}_1\eqno(2.24)
$$
We can identity $\alpha=2\lambda$ and $\beta=\lambda^2$ and then Eq. (2.24)
shows that a linear superposition of the three structures with arbitrary,
independent parameters is a Hamiltonian structure leading to compatibility.

We end this section by noting that if we define a recursion operator as
$$
R=\partial^2+{1\over 3}u+{1\over3}\partial u\partial^{-1}\eqno(2.25)
$$
then, it is easy to see that
$$
{\cal D}_3=R{\cal D}_2=R^2{\cal D}_1\eqno(2.26)
$$
This leads to the vanishing of the Nijenhuis torsion tensor associated with
$R$ which is a sufficient condition for integrability [16-19].
We note that since $R$ is a recursion operator defined from two compatible
Hamiltonian structures ${\cal D}_1$ and ${\cal D}_2$, the definition in Eq.
(2.26) would imply that ${\cal D}_3$ is Hamiltonian as well [5,18,19].
However, it
is not a priori clear that ${\cal D}_1$, ${\cal D}_2$ and ${\cal D}_3$ would
define a tri-Hamiltonian system. But we also note that under
$$
\eqalign{
u\to &\, u+{3\over2}\lambda\cr
R\to &\, R+\lambda \cr
{\cal D}_3\to &\,(R+\lambda)^2{\cal D}_1={\cal D}_3+2\lambda{\cal
D}_2+\lambda^2{\cal D}_1\cr
}\eqno(2.27)
$$
leading once again to the compatibility of the three Hamiltonian structures.

\medskip
\noindent {\bf 3. {NLS Equation as a Tri-Hamiltonian System}}
\smallskip

In this section let us consider the familiar $1+1$ dimensional system described
by
$$
\eqalign{
i{\partial q\over\partial t}&=-q_{xx}+2k (\qb q)q\cr
i{\partial \qb \over\partial t}&=\qb_{xx} -2k (\qb q)\qb \cr
}\eqno(3.1)
$$
Here $k$ is an arbitrary parameter measuring the strength of the nonlinear
interactions and can be set to unity through a rescaling of the dynamical
variables $q$ and $\qb $.

The nonlinear Schr\"odinger equation is also well known to be a bi-Hamiltonian
system [5]. Thus, for example, if we define
$$
{\bmit Q}=\pmatrix{q\cr
\noalign{\vskip 5pt}%
\qb \cr}\eqno(3.2)
$$
with
$$
\{Q_\alpha(x),Q_\beta(y)\}_1=({\cal D}_1)_{\alpha\beta}\delta(x-y)=
\pmatrix{0 & i\cr
\noalign{\vskip 5pt}%
-i & 0\cr}\delta(x-y)\qquad\alpha,\beta=1,2\eqno(3.3)
$$
and
$$
H_3=-\int dx\,\left(\qb_xq_x+k(\qb q)^2\right)\eqno(3.4)
$$
then, we obtain
$$
\eqalign{
i{\partial q\over\partial t}=&i\{q(x),H_3\}_1=-q_{xx}+2k(\qb q)q\cr
i{\partial \qb\over\partial t}=&i\{\qb(x),H_3\}_1=\qb_{xx}-2k(\qb q)\qb \cr
}\eqno(3.5)
$$
This shows that the NLS equation is a Hamiltonian system since the structure
${\cal D}_1$ in Eq. (3.3) is antisymmetric and satisfies the Jacobi identity
(trivially).

We also note that we can define
$$
\displaylines{
\{Q_\alpha(x),Q_\beta(y)\}_2=({\cal D}_2)_{\alpha\beta}\delta(x-y)\qquad\cr
\noalign{\vskip 10pt}%
\hfill=\pmatrix{kq\partial^{-1}q &{1\over2}\partial-kq\partial^{-1}\qb \cr
\noalign{\vskip 10pt}%
{1\over2}\partial-k\qb \partial^{-1}q & k\qb \partial^{-1}\qb \cr}\delta(x-y)
\hfill(3.6)\cr
}
$$
and
$$
H_2=i\int dx\,\left(\qb q_x-\qb_x q\right)\eqno(3.7)
$$
to obtain
$$
\eqalign{
i{\partial q\over\partial t}=&i\{q(x),H_2\}_2=-q_{xx}+2k(\qb q)q\cr
i{\partial \qb \over\partial t}=&i\{\qb (x),H_2\}_2=\qb_{xx}-2k(\qb q)\qb \cr
}\eqno(3.8)
$$
The second bracket structure in Eq. (3.6) is manifestly antisymmetric and is
known to satisfy the Jacobi identity. This shows that the nonlinear
Schr\"odinger equation is Hamiltonian with respect to two distinct Hamiltonian
structures. Furthermore, these two Hamiltonian structures are known to be
compatible making the nonlinear Schr\"odinger equation a bi-Hamiltonian system.

Let us next define

$$
\displaylines{
\hfill\{Q_\alpha(x),Q_\beta(y)\}_3=({\cal D}_3)_{\alpha\beta}
\delta(x-y)\hfill(3.9)\cr
\noalign{\vskip 10pt}%
\hfill=-{i\over 2}
\pmatrix{k(\partial q\partial^{-1}q-q\partial^{-1}q\partial)&
{1\over 2}\partial^2-k(\partial q\partial^{-1}\qb +q\partial^{-1}\qb
\partial)\cr
\noalign{\vskip 10pt}%
-{1\over 2}\partial^2+k(\partial \qb \partial^{-1}q+\qb
\partial^{-1}q\partial)&
-k(\partial \qb \partial^{-1}\qb -\qb \partial^{-1}\qb \partial)
\cr}\delta(x-y)\hfill\cr
}
$$
and
$$
H_1=4\int dx\, \qb q\eqno(3.10)
$$
which would give
$$
\eqalign{
i{\partial q\over\partial t}=&i\{q(x),H_1\}_3=-q_{xx}+2k(\qb q)q\cr
i{\partial \qb \over\partial t}=&i\{\qb (x),H_1\}_3=\qb_{xx}-2k(\qb q)\qb \cr
}\eqno(3.11)
$$
Therefore, if we can show that ${\cal D}_3$ defines a Hamiltonian structure, we
would have shown that the nonlinear Schr\"odinger equation is Hamiltonian with
respect to three distinct Hamiltonian structures.

To show that ${\cal D}_3$ is a Hamiltonian structure, we note from the
definition in Eq. (3.9) that it is manifestly antisymmetric. The Jacobi
identity can also be checked through the method of prolongation in the
following way. We note that the dynamical variables in the present case define
a two component vector and ${\cal D}_3$ is a $2\times2$ matrix.
Correspondingly, let us introduce
$$
{\bmit\theta}=\pmatrix{\theta\cr\thetab}
\eqno(3.12)
$$
and define a bivector as
$$
\eqalign{
{\bmit\Theta}_{{\cal D}_3}=&{1\over2}\int dx\,{\bmit\theta}^t
\wedge{\cal D}_3 {\bmit\theta}\cr
=&{i\over 2}\int dx\,
\left\{-{1\over 2}\theta_{xx}\wedge\thetab+
k(q\theta-\qb \thetab)\wedge\left(\partial^{-1}(q\theta_x+\qb \thetab_x)\right)
\right\}
}\eqno(3.13)
$$
Here ${\bmit\theta}^t$ denotes the transpose of ${\bmit\theta}$.
Once again, variables $\theta$ and $\thetab$ are assumed to be functionally
independent of $q$ and $\qb $ and prolongation acts only on functionals of $q$
and $\qb $. Thus, we obtain
$$
\displaylines{
\quad\pr\,\v_{{\cal D}_3{\bmit\theta}}({\bmit\Theta}_{{\cal D}_3})=
{ik\over 2}\int dx\,\Bigl\{(\pr\,\v_{{\cal D}_3{\bmit\theta}}(q)
\wedge\theta-\pr\,\v_{{\cal D}_3{\bmit\theta}}(\qb )\wedge\thetab)
\wedge\left(\partial^{-1}(q\theta_x+\qb \thetab_x)\right)\hfill\cr
\hfill-(q\theta-\qb \thetab)\wedge(\partial^{-1}
(\pr\,\v_{{\cal D}_3{\bmit\theta}}(q)\wedge\theta_x+
\pr\,\v_{{\cal D}_3{\bmit\theta}}(\qb )\wedge\thetab_x)
\Bigr\}\quad(3.14)
}
$$
Using the relations for the present case,
$$
\eqalign{
\pr\,\v_{{\cal D}_3{\bmit\theta}}(q)=&
-{i\over4}\thetab_{xx}
+{ik\over2}q\left(\partial^{-1}(q\theta_x+\qb\thetab_x)\right)
-{ik\over2}\left(q\left(\partial^{-1}(q\theta-\qb\thetab)\right)\right)_x\cr
\pr\,\v_{{\cal D}_3{\bmit\theta}}(\qb )=&
{i\over4}\theta_{xx}
-{ik\over2}\qb\left(\partial^{-1}(q\theta_x+\qb\thetab_x)\right)
-{ik\over2}\left(\qb\left(\partial^{-1}(q\theta-\qb\thetab)\right)\right)_x
}\eqno(3.15)
$$
it is straightforward to check that
$$
\pr\,\v_{{\cal D}_3{\bmit\theta}}({\bmit\Theta}_{{\cal
D}_3})=0\eqno(3.16)
$$
This shows that the structure, ${\cal D}_3$, defines a Hamiltonian structure.

To show compatibility of the three Hamiltonian structures, we define as before
$$
{\cal D}={\cal D}_3+\alpha{\cal D}_2+\beta{\cal D}_1\eqno(3.17)
$$
where $\alpha$ and $\beta$ are arbitrary, independent constants. By
construction, ${\cal D}$ is antisymmetric since each of the Hamiltonian
structures ${\cal D}_1$, ${\cal D}_2$ and ${\cal D}_3$ is. To check Jacobi
identity, we again construct a bivector
$$
{\bmit\Theta}_{{\cal D}}={1\over2}\int dx\, {\bmit\theta}^t
\wedge{\cal D} {\bmit\theta}
={1\over 2}\int dx\,\Bigl\{{\bmit\theta}^t\wedge
({\cal D}_3{\bmit\theta}+\alpha{\cal D}_2{\bmit\theta}
+\beta{\cal D}_1{\bmit\theta})\Bigr\}\eqno(3.18)
$$
It is, then, tedious but straightforward to check that
$$
\pr\,\v_{{\cal D}{\bmit\theta}}({\bmit\Theta}_{{\cal
D}})=0\eqno(3.19)
$$
This shows that the three Hamiltonian structures ${\cal D}_1$, ${\cal D}_2$ and
${\cal D}_3$ are compatible making the nonlinear Schr\"odinger equation a
tri-Hamiltonian system.

We end this section by noting that if we define a matrix recursion operator as
$$
R=
\pmatrix{-{i\over2}\partial+ikq\partial^{-1}\qb &ik q\partial^{-1}q \cr
\noalign{\vskip 10pt}%
-ik\qb \partial^{-1}\qb & {i\over2}\partial-ik\qb \partial^{-1}q\cr}
\eqno(3.20)
$$
then, we can write
$$
{\cal D}_3=R{\cal D}_2=R^2{\cal D}_1\eqno(3.21)
$$
Once again, this would imply that the Nijenhuis torsion tensor associated with
$R$ vanishes which is a sufficient condition for integrability [16-19]. Once
again since $R$ is constructed from two compatible structures ${\cal D}_1$ and
${\cal D}_2$, it would also imply that ${\cal D}_3$ is Hamiltonian.
We also note that
$R\to R+\lambda I$ would provide an alternate way of looking at the
compatibility of these structures. However, we have not succeeded in finding a
transformation of the dynamical variables which will generate this shift in the
recursion operator.
\medskip
\noindent {\bf 4. {Two Boson Hierarchy}}
\smallskip

It is known that the two boson hierarchy equation [6]
$$
\eqalign{
{\partial u\over\partial t}=&(2h+u^2-u_x)_x\cr
{\partial h\over\partial t}=&(2uh+h_x)_x\cr
}\eqno(4.1)
$$
yields the nonlinear Schr\"odinger equation (we will assume $k=1$)  with the
field redefinitions [7-10]
$$
\eqalign{
u=&-{q_x\over q}\cr
h=&-\qb q
}\eqno(4.2)
$$
and the coordinate scaling
$$
t\to it\eqno(4.3)
$$
The converse is also true, namely, we can obtain the two boson hierarchy from
the nonlinear Schr\"odinger equation through an inverse field redefinition and
coordinate transformation. In this section, we will show how we can obtain the
three Hamiltonian structures of the two boson hierarchy starting from the
structures of the nonlinear Schr\"odinger equation described in the previous
section.

To that end, we define
$$
{\bmit U}=\pmatrix{u\cr
\noalign{\vskip 5pt}%
h\cr}\eqno(4.4)
$$
and note that with the relations in Eq. (4.2) and the definition in Eq. (3.2),
we can think of $\bmit U$ as a functional of $\bmit Q$, namely,
${\bmit U}[{\bmit Q}]$. Let
us next define (We use an abstract operator notation for simplicity.
Coordinates can be brought in by taking appropriate matrix elements.)
$$
P_{\alpha\beta}={\delta U_\alpha\over\delta Q_\beta}=
\pmatrix{-\partial{\rm e}^{(\partial^{-1}u)}& 0\cr
\noalign{\vskip 5pt}%
h\,{\rm e}^{(\partial^{-1}u)} & -{\rm e}^{-(\partial^{-1}u)}\cr}
\eqno(4.5)
$$
We also define the formal adjoint [20] of $P$ as
$$
P^*_{\alpha\beta}=
\pmatrix{{\rm e}^{(\partial^{-1}u)}\partial&h\,{\rm e}^{(\partial^{-1}u)} \cr
\noalign{\vskip 5pt}%
0 & -{\rm e}^{-(\partial^{-1}u)}\cr}
\eqno(4.6)
$$
If we now denote the $2\times2$ matrix Hamiltonian structure of $U_\alpha$ as
$$
\{U_\alpha,U_\beta\}={\widetilde D}_{\alpha\beta}\eqno(4.7)
$$
then, it is easy to see that
$$
{\widetilde D}=P{\cal D}P^*\eqno(4.8)
$$
where $\cal D$ is the corresponding Hamiltonian structure for the $Q_\alpha$'s.

With the three Hamiltonian structures for the nonlinear Schr\"odinger equation
defined in Eqs. (3.3), (3.6) and (3.9) and the matrices $P$ and $P^*$ in Eqs.
(4.5) and (4.6), it can now be easily checked that
\eject
$$
\eqalignno{
{\widetilde D}_1=&P{\cal D}_1P^*=i\pmatrix{0 &\partial \cr
\noalign{\vskip 5pt}%
\partial & 0\cr}=iD_1 &(4.9)\cr
\noalign{\vskip 10pt}%
{\widetilde D}_2=&P{\cal D}_2P^*=-{1\over2}
\pmatrix{2\partial &\partial(u-\partial) \cr
\noalign{\vskip 5pt}%
(\partial+u)\partial &(\partial h+h\partial) \cr}=-{1\over2}D_2 &(4.10)\cr
\noalign{\vskip 10pt}%
{\widetilde D}_3=&P{\cal D}_3P^*\cr
=&-{i\over2}\pmatrix{(\partial u+ u\partial) & (\partial
h+h\partial)+{1\over2}\partial(\partial-u)^2\cr
\noalign{\vskip 5pt}%
(\partial h+h\partial)+{1\over2}(\partial+u)^2\partial &
{1\over2}(\partial+u)(\partial h+h\partial)+{1\over2}(\partial
h+h\partial)(u-\partial)\cr}\cr
=&-{i\over 4}D_3 &(4.11)\cr
}
$$
where $D_1$, $D_2$ and $D_3$ are the three Hamiltonian structures for the two
boson hierarchy as given in ref. 6 (The constant multiples can always be
defined away by rescaling the corresponding Hamiltonians.) We note here that
the compatibility of the Hamiltonian structures $D_1$, $D_2$ and
$D_3$ can be easily seen by shifting $u\to u+\lambda$ in the context of
the two boson theory.

\medskip
\noindent {\bf 5. {Conclusion}}
\smallskip

We have derived a third Hamiltonian structure for the KdV equation and have
shown that all three structures are compatible making KdV a tri-Hamiltonian
system. We have also derived a third Hamiltonian structure for the NLS equation
and have shown that it is tri-Hamiltonian as well. The proof of Jacobi identity
and the compatibility are carried out through the method of prolongation and we
have commented on the properties of the recursion operators for these system.
We have also shown how the three Hamiltonian structures for the two boson
hierarchy can be obtained from those for the NLS equation. We speculate, based
on our results, that integrable systems where the Hamiltonian description of
the dynamical equations has not exhausted the lowest possible local
Hamiltonian,
are likely to be tri-Hamiltonian systems.
\medskip
\noindent {\bf Acknowledgements}
\smallskip

This work was supported in part by the U.S. Department of Energy Grant No.
DE-FG-02-91ER40685. J.C.B. would like to thank CNPq, Brazil, for
financial support.
\vfill\eject
\noindent {\bf {Appendix}}
\smallskip

For completeness we list here all the prolongation formulae used in the text.
\bigskip
\noindent {\it Prolongation Formulae for {\rm KdV}:}
\smallskip

For all structures $\cal D$ we have
$$
\pr\,\v_{{\cal D}\theta}(u)={\cal D}\theta
$$
\noindent The prolongations formulae for ${\cal D}_1$ are
$$
\eqalign{
{\cal D}_1\theta=&\theta_x\cr
{\bmit\Theta}_{{\cal D}_1}=&{1\over2}\int dx\,{\theta}\wedge{\theta}_x \cr
\pr\,\v_{{\cal D}_1{\theta}}({\bmit\Theta}_{{\cal D}_1})=&0\hbox{ (trivially)}
}
$$
For ${\cal D}_2$ they are
$$
\eqalign{
{\cal D}_2\theta=&\theta_{xxx}+{1\over3}(u\theta)_x+{1\over3}u\theta_x\cr
{\bmit\Theta}_{{\cal D}_2}=&{1\over2}\int dx\,\left\{
{\theta}\wedge{\theta}_{xxx}+{2\over3}u\theta\wedge\theta_x\right\} \cr
\pr\,\v_{{\cal D}_2{\theta}}({\bmit\Theta}_{{\cal D}_2})=&{1\over3}\int dx\,
\left\{\pr\,\v_{{\cal D}_2{\theta}}(u)\wedge\theta\wedge\theta_x\right\}=0
}
$$
And finally for ${\cal D}_3$ we have
$$
\eqalign{
{\cal D}_3\theta=&\theta_{xxxxx}+{1\over3}(u\theta)_{xxx}+
{1\over3}(u\theta_x)_{xx}+{1\over3}(u\theta_{xx})_{x}
+{1\over3}u\theta_{xxx}\cr
&\phantom{\theta_{xxxxx}}
+{1\over9}(u^2\theta)_x +{1\over9}u(u\theta)_x+{1\over9}u^2\theta_x
+{1\over9}\left(u(\partial^{-1}u\theta_x)\right)_x\cr
{\bmit\Theta}_{{\cal D}_3}=&{1\over2}\int dx\,\Biggl\{
{\theta}\wedge{\theta}_{xxxxx}+{2\over3}u\theta\wedge\theta_{xxx}
-{2\over 3}u\theta_x\wedge\theta_{xx}\cr
&\phantom{{1\over2}\int dx\,\Biggl\{{\theta}\wedge{\theta}_{xxxxx}}
+{1\over 3}u^2\theta\wedge\theta_{x}-
{1\over9}u\theta_x\wedge(\partial^{-1}u\theta_x)\Biggr\}\cr
\pr\,\v_{{\cal D}_3{\theta}}({\bmit\Theta}_{{\cal D}_3})=&
{1\over2}\int dx\,\Biggl\{{2\over3}\pr\,\v_{{\cal D}_3{\theta}}(u)
\wedge(\theta\wedge\theta_{xxx}-\theta_x\wedge\theta_{xx}
+u\theta\wedge\theta_x)\cr
-&{1\over 9} \pr\,\v_{{\cal D}_3{\theta}}(u)\wedge\theta_x
\wedge(\partial^{-1}u\theta_x)
+{1\over 9}u\theta_x\wedge\left(\partial^{-1}
\pr\,\v_{{\cal D}_3{\theta}}(u)\wedge\theta_x\right)\Biggr\}=0\cr
}
$$
\vfill\eject
\noindent For the compatibility of the structures, we have
$$
\eqalign{
{\cal D}=&{\cal D}_3+\alpha{\cal D}_2+\beta{\cal D}_1\cr
{\cal D}\theta=&{\cal D}_3\theta+\alpha{\cal D}_2\theta+\beta{\cal
D}_1\theta\cr
\pr\,\v_{{\cal D}{\theta}}({\bmit\Theta}_{{\cal D}})=&
\pr\,\v_{{\cal D}{\theta}}({\bmit\Theta}_{{\cal D}_3})+
\alpha\pr\,\v_{{\cal D}{\theta}}({\bmit\Theta}_{{\cal D}_2})+
\beta\pr\,\v_{{\cal D}{\theta}}({\bmit\Theta}_{{\cal D}_1})=0
}
$$
\bigskip
\noindent {\it Prolongation Formulae for {\rm NLS}:}
\smallskip

In this case,
$\bmit\theta$, $\bmit Q$ and ${\cal D}{\bmit\theta}$ are  two-component vectors
and
$$
\pr\,\v_{{\cal D}{\bmit\theta}}(\bmit Q)={\cal D}{\bmit\theta}
$$
for any ${\cal D}$. $\pr\,\v_{{\cal D}{\bmit\theta}}(q)$ and
$\pr\,\v_{{\cal D}{\bmit\theta}}(\qb)$, then, would correspond to the first and
the second components of ${\cal D}{\bmit\theta}$.

The prolongations formulae for ${\cal D}_1$ are
$$
\eqalign{
{\cal D}_1{\bmit\theta}=&i\pmatrix{\thetab\cr-\theta}\cr
{\bmit\Theta}_{{\cal D}_1}=&i\int dx\,{\theta}\wedge{\thetab}\cr
\pr\,\v_{{\cal D}_1{\bmit\theta}}({\bmit\Theta}_{{\cal D}_1})=&0
}
$$
For ${\cal D}_2$ they are
$$
\eqalign{
{\cal D}_2{\bmit\theta}=&\pmatrix{
{1\over2}\thetab_x+{k\over2}q\left(\partial^{-1}(q\theta-\qb\thetab)\right)\cr
\noalign{\vskip 10pt}%
{1\over2}\theta_x-{k\over2}\qb\left(\partial^{-1}(q\theta-\qb\thetab)\right)\cr
}\cr
\noalign{\vskip 15pt}%
{\bmit\Theta}_{{\cal D}_2}=&{1\over2}\int dx\,
\left\{{\theta}\wedge\thetab_x+{k\over2}(q\theta-\qb\thetab)\wedge
\left(\partial^{-1}(q\theta-\qb\thetab)\right)\right\}\cr
\noalign{\vskip 15pt}%
\pr\,\v_{{\cal D}_2{\bmit\theta}}({\bmit\Theta}_{{\cal D}_2})=&{k\over2}\int
dx\,\Bigl\{
\left(\pr\,\v_{{\cal D}_2{\bmit\theta}}(q)\wedge\theta-
\pr\,\v_{{\cal D}_2{\bmit\theta}}(\qb)\wedge\thetab\right)\wedge
\left(\partial^{-1}(q\theta-\qb\thetab)\right)\Bigr\}=0
}
$$
\vfill\eject
\noindent Finally, for ${\cal D}_3$ we have
$$
\eqalign{
{\cal D}_3{\bmit\theta}=&\pmatrix{
-{i\over4}\thetab_{xx}
+{ik\over2}q\left(\partial^{-1}(q\theta_x+\qb\thetab_x)\right)
-{ik\over2}\left(q\left(\partial^{-1}(q\theta-\qb\thetab)\right)\right)_x\cr
\noalign{\vskip 10pt}%
{i\over4}\theta_{xx}
-{ik\over2}\qb\left(\partial^{-1}(q\theta_x+\qb\thetab_x)\right)
-{ik\over2}\left(\qb\left(\partial^{-1}(q\theta-\qb\thetab)\right)\right)_x
}\cr
\noalign{\vskip 15pt}%
{\bmit\Theta}_{{\cal D}_3}=&-{i\over4}\int dx\,
\Bigl\{{\theta}\wedge\thetab_{xx}-2k(q\theta-\qb\thetab)\wedge
\left(\partial^{-1}(q\theta_x+\qb\thetab_x)\right)\Bigr\}\cr
\noalign{\vskip 15pt}%
\pr\,\v_{{\cal D}_3{\bmit\theta}}({\bmit\Theta}_{{\cal D}_3})
=&{ik\over2}\int dx\,\Bigl\{
\left(\pr\,\v_{{\cal D}_3{\bmit\theta}}(q)\wedge\theta-
\pr\,\v_{{\cal D}_3{\bmit\theta}}(\qb)\wedge\thetab\right)\wedge
\left(\partial^{-1}(q\theta_x+\qb\thetab_x)\right)\cr
&\phantom{{ik\over2}\int dx\,}-(q\theta-\qb\thetab)\wedge
\left(\partial^{-1}\left(\pr\,\v_{{\cal D}_3{\bmit\theta}}(q)\wedge\theta_x+
\pr\,\v_{{\cal D}_3{\bmit\theta}}(\qb)\wedge\thetab_x\right)\right)\Bigr\}=0
}
$$
For compatibility, we have
$$
\eqalign{
{\cal D}=&{\cal D}_3+\alpha{\cal D}_2+\beta{\cal D}_1\cr
{\cal D}{\bmit\theta}=&{\cal D}_3{\bmit\theta}+\alpha{\cal D}_2{\bmit\theta}+
\beta{\cal D}_1{\bmit\theta}\cr
\pr\,\v_{{\cal D}{\bmit\theta}}({\bmit\Theta}_{{\cal D}})=&
\pr\,\v_{{\cal D}{\bmit\theta}}({\bmit\Theta}_{{\cal D}_3})+
\alpha\pr\,\v_{{\cal D}{\bmit\theta}}({\bmit\Theta}_{{\cal D}_2})+
\beta\pr\,\v_{{\cal D}{\bmit\theta}}({\bmit\Theta}_{{\cal D}_1})=0
}
$$

\vfill\eject

\noindent {\bf {References}}
\medskip

\item{1.} L.D. Faddeev and L.A. Takhtajan, ``Hamiltonian Methods in
the Theory of Solitons'' (Springer, Berlin, 1987).

\item{2.} A. Das, ``Integrable Models'' (World Scientific, Singapore,
1989).

\item{3.} M.J. Ablowitz and P.A. Clarkson, ``Solitons, Nonlinear
Evolution Equations and Inverse Scattering'' (Cambridge, New York, 1991).

\item{4.} L. A. Dickey, ``Soliton Equations and Hamiltonian Systems'' (World
Scientific, Singapore, 1991).

\item{5.} F. Magri, J. Math. Phys. {\bf 19} (1978) 1156.

\item{6.} B.A. Kupershmidt, Commun. Math. Phys. {\bf 99} (1985) 51.

\item{7.} H. Aratyn, L.A. Ferreira, J.F. Gomes and A.H. Zimerman, Nucl.
 Phys. {\bf B402} (1993) 85; H. Aratyn, L.A. Ferreira, J.F. Gomes and A.H.
Zimerman, ``Lectures at the VII J. A. Swieca Summer School'', January 1993,
hep-th/9304152; H. Aratyn, E. Nissimov and S. Pacheva, Phys. Lett. {\bf B314}
(1993) 41.

\item{8.} L. Bonora and C.S. Xiong, Phys. Lett. {\bf B285} (1992) 191; L.
Bonora and C.S. Xiong, Int. J. Mod. Phys. {\bf A8} (1993) 2973.

\item{9.} M. Freeman and P. West, Phys. Lett. {\bf 295B} (1992) 59.

\item{10.} J. Schiff, ``The Nonlinear Schr\"odinger Equation and
Conserved Quantities in the Deformed Parafermion and SL(2,{\bf R})/U(1)
Coset Models'', Princeton preprint IASSNS-HEP-92/57 (1992)
(also hep-th/9210029).

\item{11.} J. C. Brunelli, A. Das and W.-J. Huang, Mod. Phys. Lett. {\bf 9A}
(1994) 2147.

\item{12.} W. Oevel and W. Strampp, Commun. Math. Phys. {\bf 157} (1993) 51.

\item{13.} We note that there exist other Hamiltonian systems with three or
more Hamiltonian structures which are only pairwise compatible. For example,
see P. J. Olver and Y. Nutku, J. Math. Phys. {\bf 29} (1988) 1610.

\item{14.} P. J. Olver , ``Applications of Lie Groups to Differential
Equations'', Graduate Texts in Mathematics, Vol. 107 (Springer, New York,
1986).

\item{15.} J. L. Gervais, Phys. Lett. {\bf B160} (1985) 277.

\item{16.} S. Okubo and A. Das, Phys. Lett. {\bf 209B} (1988) 311;
A. Das and S. Okubo, Ann. Phys. {\bf 190} (1989) 215.

\item{17.} C. Ferrario, G. LoVecchio, G. Marmo and G. Morandi, Lett. Math.
Phys. {\bf 9} (1985) 140; P. Antonini, G. Marmo and C. Rubano, Nuovo Cimento
{B86} (1985) 17.

\item{18.} S. Okubo, J. Math. Phys. {\bf 30} (1989) 834; S. Okubo, Alg. Group.
Geom. {\bf 6} (1989) 65.

\item{19.} A. Das and W.-J. Huang, J. Math. Phys. {\bf 31} (1990) 2603.

\item{20.} E. Date, M. Kashiwara, M. Jimbo and T. Miwa, in ``Nonlinear
Integrable Systems - Classical Theory and Quantum Theory'', ed. M. Jimbo and T.
Miwa (World Scientific, Singapore, 1983).

\end